\documentclass{elsart3p}

\usepackage{ifpdf}
\usepackage{graphicx,amssymb,lineno}
\ifpdf
\usepackage[%
  pdftitle={Instructions for use of the document class
    elsart},%
  pdfauthor={Simon Pepping},%
  pdfsubject={The preprint document class elsart},%
  pdfkeywords={instructions for use, elsart, document class},%
  pdfstartview=FitH,%
  bookmarks=true,%
  bookmarksopen=true,%
  breaklinks=true,%
  colorlinks=true,%
  linkcolor=blue,anchorcolor=blue,%
  citecolor=blue,filecolor=blue,%
  menucolor=blue,pagecolor=blue,%
  urlcolor=blue]{hyperref}
\else
\usepackage[%
  breaklinks=true,%
  colorlinks=true,%
  linkcolor=blue,anchorcolor=blue,%
  citecolor=blue,filecolor=blue,%
  menucolor=blue,pagecolor=blue,%
  urlcolor=blue]{hyperref}
\fi

\usepackage{graphics}
\usepackage{epsfig}

\makeatletter
\def\elsartstyle{%
    \def\normalsize{\@setfontsize\normalsize\@xiipt{14.5}}
    \def\small{\@setfontsize\small\@xipt{13.6}}
    \let\footnotesize=\small
    \def\large{\@setfontsize\large\@xivpt{18}}
    \def\Large{\@setfontsize\Large\@xviipt{22}}
    \skip\@mpfootins = 18\p@ \@plus 2\p@
    \normalsize
}
\@ifundefined{square}{}{\let\Box\square}
\makeatother

\def\file#1{\texttt{#1}}

\newif\ifdummy\dummyfalse

\begin{document}

\begin{frontmatter}



\vspace{-2.cm}
\title{
The BAIKAL neutrino experiment $–-$ 
physics results and perspectives
}

\author[a]{V. Aynutdinov},
\author[a]{A.Avrorin},
\author[a]{V. Balkanov},
\author[d]{I. Belolaptikov},
\author[b]{D. Bogorodsky},
\author[b]{N. Budnev},
\author[a]{I. Danilchenko},
\author[a]{G. Domogatsky},
\author[a]{A. Doroshenko},
\author[b]{A. Dyachok},
\author[a]{Zh.-A. Dzhilkibaev},
\author[f]{S. Fialkovsky},
\author[a]{O. Gaponenko},
\author[d]{K.Golubkov},
\author[b]{O. Gress},
\author[b]{T. Gress},
\author[b]{O. Grishin},
\author[a]{A. Klabukov},
\author[h]{A. Klimov},
\author[b]{A.Kochanov},
\author[d]{K. Konischev},
\author[a]{A. Koshechkin},
\author[f]{V. Kulepov},
\author[a]{D. Kuleshov},
\author[c]{L. Kuzmichev},
\author[b]{S. Lovtsov},
\author[e]{E. Middell},
\author[a]{S. Mikheyev},
\author[f]{M. Milenin},
\author[b]{R. Mirgazov},
\author[c]{E. Osipova},
\author[b]{G. Pan'kov},
\author[b]{L. Pan'kov},
\author[a]{A. Panfilov},
\author[a]{D. Petukhov},
\author[d]{E. Pliskovsky},
\author[a]{P. Pokhil},
\author[a]{V. Poleschuk},
\author[c]{E. Popova},
\author[b]{A. Rastegin},
\author[c]{V. Prosin},
\author[g]{M. Rozanov},
\author[b]{V. Rubtzov},
\author[a]{A. Sheifler},
\author[c]{A. Shirokov},
\author[d]{B. Shoibonov},
\author[e]{Ch. Spiering},
\author[a]{O. Suvorova},
\author[b]{B. Tarashansky},
\author[e]{R. Wischnewski\corauthref{cor}},
\corauth[cor]{Corresponding author.Tel: +49 3376277348, Fax: +49 3376277330}
\ead{wischnew@ifh.de}
\author[c]{I. Yashin},
\author[a]{V. Zhukov}

\address[a]{Institute for Nuclear Research, 60th October Anniversary pr. 7a, 
Moscow 117312, Russia}
\address[b]{Irkutsk State University, Irkutsk, Russia}
\address[c]{Skobeltsyn Institute of Nuclear Physics  MSU, Moscow, Russia}
\address[d]{Joint Institute for Nuclear Research, Dubna, Russia}
\address[e]{DESY, Zeuthen, Germany}
\address[f]{Nizhni Novgorod State Technical University, Nizhni Novgorod, 
Russia}
\address[g]{St.Petersburg State Marine University, St.Petersburg, Russia}
\address[h]{Kurchatov Institute, Moscow, Russia}

\begin{abstract}
We review the status of the Lake Baikal Neutrino Experiment. 
The Neutrino Telescope NT200 has been operating 
since 1998 and has been upgraded to the $10$ Mton detector 
NT200+ in 2005. 
We present selected astroparticle physics results from long-term operation of NT200.
Also discussed are activities towards
acoustic detection of UHE-energy neutrinos, 
and results of associated science activities.
Preparation towards 
a km3-scale (Gigaton volume) detector in Lake Baikal is currently a central activity. 
As an important milestone,
a km3-prototype string,
based on completely new technology,
has been installed and is operating together with NT200+ since April, 2008. 
\end{abstract}

\end{frontmatter}

\section{Introduction}
\label{intro}
\begin{figure}[htb]
\centering
\includegraphics[width=0.45\textwidth]{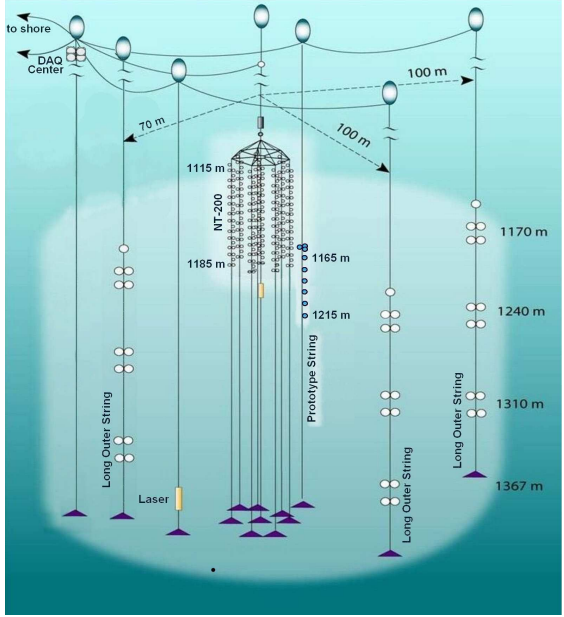}
\caption{
  The Baikal Telescope NT200+ as of 2008: the compact NT200 (center), 
  3 long outer strings and the new technology km3-prototype string.
}
\label{fig_nt200p}
\end{figure}

The Baikal Neutrino Telescope NT200 has been taking data since 1998;
its first stage telescope NT36 was - back in 1993 - the 
first underwater Cherenkov 
neutrino detector. 
Since 2005, the upgraded $10$-Mton scale detector NT200$+$ 
is in operation.
Detector configuration and performance have been 
described elsewhere \cite{APP1,NE05,HEB,NANP05,RW,ICRCS}. 
The most recent milestone of the ongoing km3-telescope 
research and development work (R\&D) was the installation 
of a ``new technology'' prototype string in spring 2008,
operating now as part of NT200$+$ \cite{Toulon_string}.
Fig.\ref{fig_nt200p} gives a sketch of the current status of the 
telescope NT200+, 
including the km3-prototype string.

In this paper we review 
astroparticle physics results obtained with NT200,
a feasibility study on acoustic neutrino detection,
and the R\&D
activities towards a km3-scale Baikal 
telescope.

The success of the Baikal neutrino experiment is,
to a considerable degree, 
related 
to the favorable natural conditions of the site -- 
thus site studies and related science 
were always an integral part of the project.
We will discuss  some of the 
relevant problems for the lake's ecosystem, 
which were studied through the use of new technologies, 
instruments and methods designed in the framework 
of the Baikal experiment. 
We will also mention future options.

\section{Selected physics results from NT200}
\label{results}

\subsection{Atmospheric neutrinos}
\label{sect_munu}

\begin{figure}[t]
\includegraphics[width=0.45\textwidth, trim= 0cm 2.5cm 1cm 2.5cm]{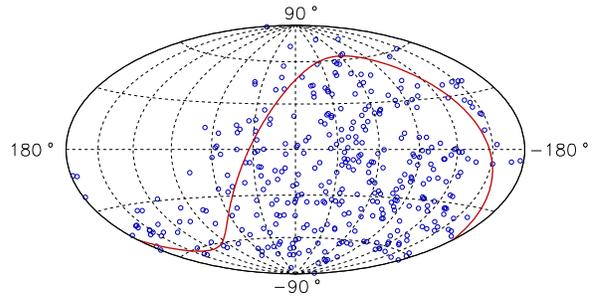}
\caption{
  Skyplot (galactic coordinates) of neutrino events for five years. 
  The solid curve shows the equator.
}
\label{fig_skyplot}
\end{figure}

The signature of charged current muon neutrino events is a 
muon crossing the detector from below. Muon track reconstruction 
algorithms and background rejection have been described elsewhere \cite{APP2}. 
Compared to \cite{APP2}, the analysis of the $5$-year sample 
(April 1998 -- February 2003, $1008$\,days live time) was optimized for 
higher signal passing rate, 
i.e. accepting a slightly higher contamination of $\sim$$20$\% 
fake events \cite{ECRS04}.
A total of $372$ upward going neutrino candidates were selected. 
From Monte-Carlo simulation a total of $385$ atmospheric neutrino 
and background events are expected,
with a median muon-angular resolution of 2.2$^o$. 
The skyplot for this event sample is shown in Fig.\ref{fig_skyplot}.

\begin{figure}[b]
\begin{center}
\hspace*{-0.5cm}
\begin{minipage}[c]{0.5\linewidth}
\hspace*{-0.5cm}
\includegraphics[width=1.1\linewidth]{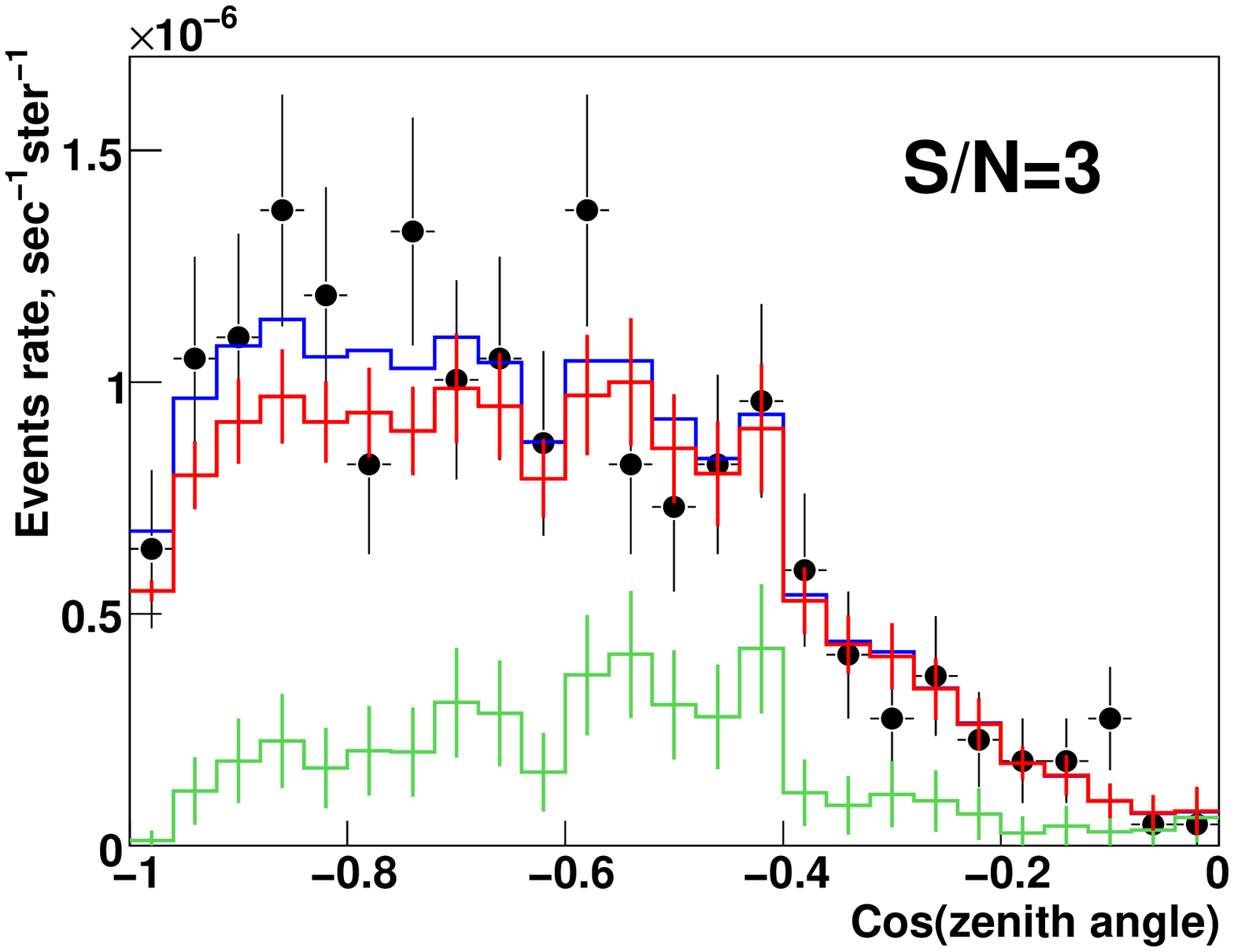}
\hspace*{-0.5cm}
\includegraphics[width=1.1\linewidth]{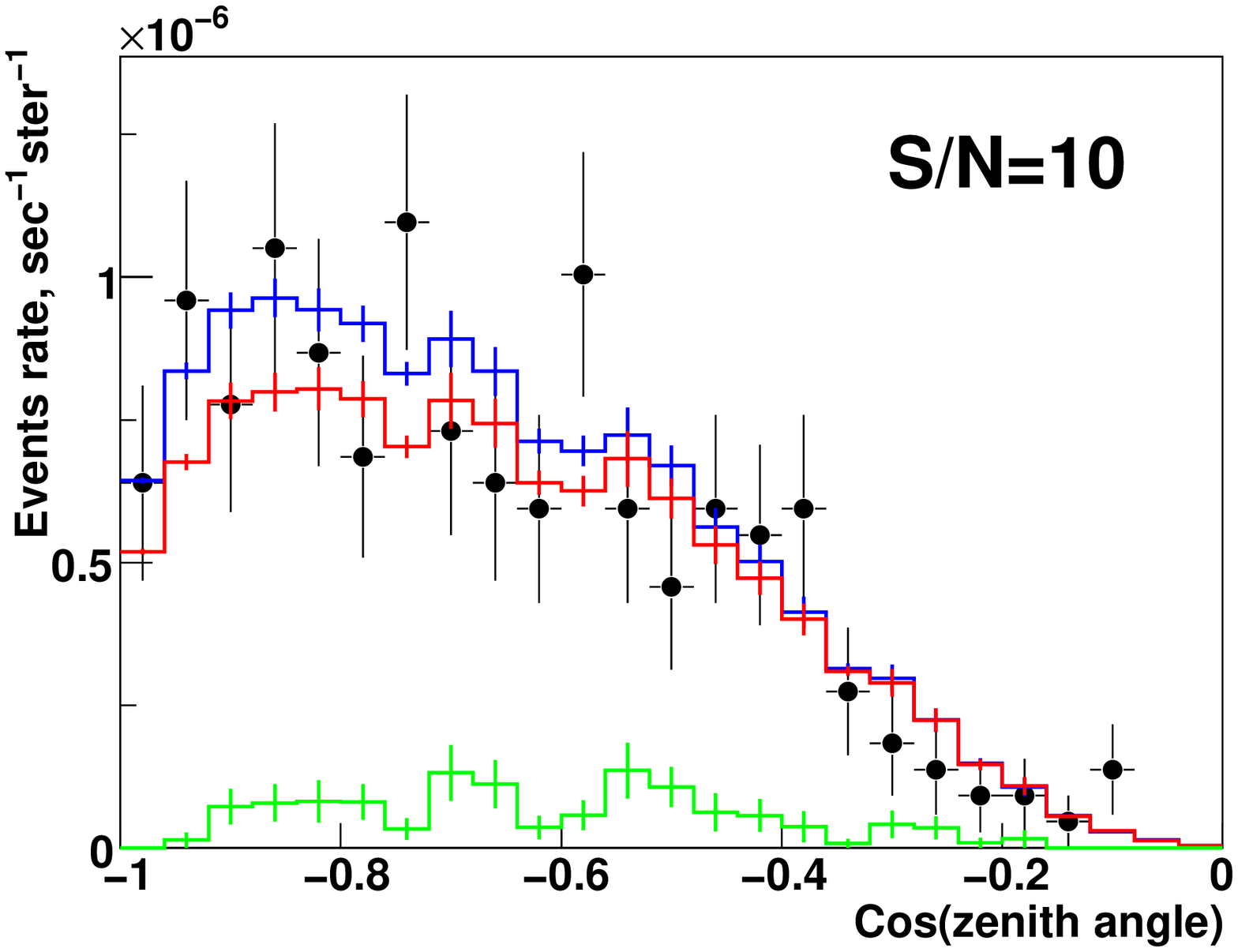}
\end{minipage}
\hspace*{-.4cm}
\begin{minipage}[c]{0.5\linewidth}
\begin{center}
\includegraphics[width=1.17\linewidth]{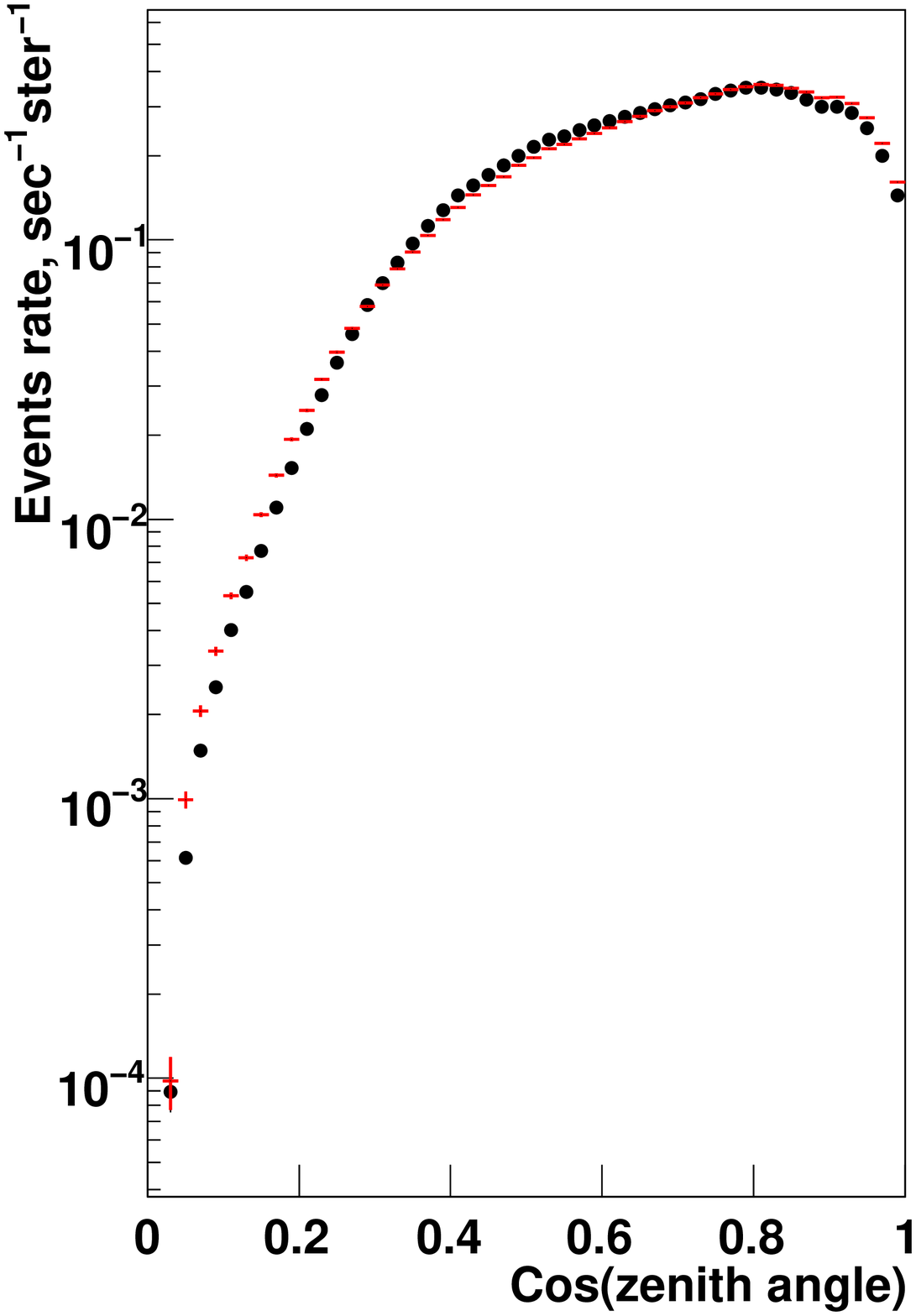}
\end{center}
\end{minipage}
\caption
{Distribution of cos(zenith) for muon events.
Left: neutrino event samples 
(data - symbols, MC - histograms (from top): sig+bkg for non-osc., 
oscillation and bkg);
Right: downward atmopsheric muons (data - symbols, MC - histogram).}
\label{fig_costheta}
\end{center}
\end{figure}

For the $\nu_{\mu}$-analysis procedure,
a standard 3-dimensional muon reconstruction is performed 
for all events \cite{APP2}.
The good agreement of the 
obtained
downward muon angular distribution with MC
is shown in
Fig.\ref{fig_costheta}(right) 
(see also earlier Baikal flux measurements \cite{APP1}).
All events reconstructed as upgoing are treated in a second pass
as neutrino candidates, to additionally filter out fake events. 
The zenith-angular distribution of upward reconstructed 
neutrino events is given in Fig.\ref{fig_costheta}(left),
for two different choices 
of the background content (S/N$\sim$3 and $\sim$10,
respectively), as used for different analyses.
In Fig.\ref{fig_costheta} also shown are MC predictions 
for the sum of $\nu$-signal and background
without and 
with $\nu$-oscillations (see Sect.\ref{sect_wimp} for parameters),
as well as for the background
(histograms from top to bottom, respectively).
Data and MC 
are in good agreement, given the known systematic uncertainties 
of absolute atmospheric $\nu$-flux predictions.

\subsection{Search for Neutrinos from WIMP Annihilation}
\label{sect_wimp}
The search for WIMPs annihilating in the Earth center with the 
Baikal neutrino telescope is based on a possible signal of
nearly vertically upward going muons, exceeding
the flux of atmospheric neutrinos. 
Signal event selection relies on 
a series of cuts which are tailored to the response
of the telescope to nearly vertically upward moving muons 
\cite{Bal1}.
These cuts select muons with $-1<\cos(\theta)<-0.75$ 
and result in a detection area of about $1800$ m$^2$ for vertically 
upward going muons.

\begin{figure}[htbp]
\centering
\includegraphics[width=0.35\textwidth, trim= 0cm 1cm 0cm 0cm]{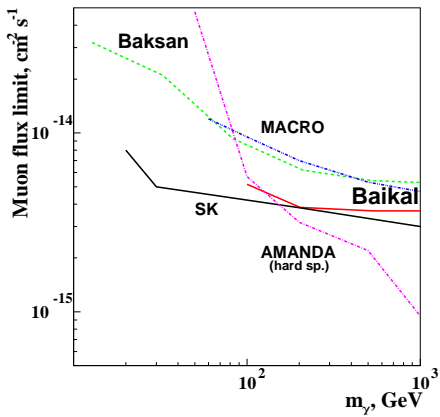}
\parbox[t]{0.47\textwidth}
{\vspace*{-0.2cm}
\caption
{Limits on the excess of muon flux from the center of 
the Earth as a function of neutralino mass.}
\label{fig_wimpmass}}
\hfill
\end{figure}

The energy threshold for this analysis is $E_{\mbox{thr}}\sim 10$ GeV
i.e. 
lower then for the 
standard $\nu_{\mu}$-analysis 
described in 
Sect.\ref{sect_munu}
($E_{\mbox{thr}}\sim 15 - 20$ GeV).

From 1038 live time days 
(1998-2003),
48 events with -1$< \cos(\theta) <$-0.75 have been selected 
as clear neutrino events, compared to 56.6 events expected
from an atmospheric neutrino MC 
(Bartol-96 flux \cite{Bartol} with 
oscillation parameters $\delta m^2 = $2.5$\cdot$10$^{-3}$ eV$^2$ 
and full mixing, $\theta_m\approx \pi/$4;
without  
oscillation 73.1 MC-events are expected);
for details see \cite{WIMP05,RICAP07}.
We find that absolute number and zenith angular distributions 
for MC  and experimental data 
are within statistical uncertainties 
in good agreement.

Regarding the 48 detected events as being induced by atmospheric 
neutrinos, one can derive an upper limit on the additional flux 
of muons from the center of the Earth   
due to annihilation of neutralinos - the favored candidate for
cold dark matter.
From this, we obtain 90\% C.L. muon flux limits for six cones 
around the opposite zenith 
($E_{\mbox{thr}}>$10 GeV).
From the strong dependence of the size of the cone 
on the neutralino mass, see \cite{BAKSANWIMP, MACROWIMP, SKWIMP}, 
we calculate 
90\% C.L. flux limits 
as function of neutralino
mass.
A correction 
is applied for each neutralino mass to translate
from the experimental $10$ GeV to 
$1$ GeV threshold. 
These limits are shown in Fig.\ref{fig_wimpmass}. 
Also shown 
are limits
obtained by Baksan \cite{BAKSANWIMP}, 
MACRO \cite{MACROWIMP}, Super-Kamiokande \cite{SKWIMP} and AMANDA 
(from the hard neutralino annihilation channels)\cite{AMANDAWIMP}.

\begin{figure}[htbp]
\includegraphics[width=0.35\textwidth, height=0.35\textwidth]{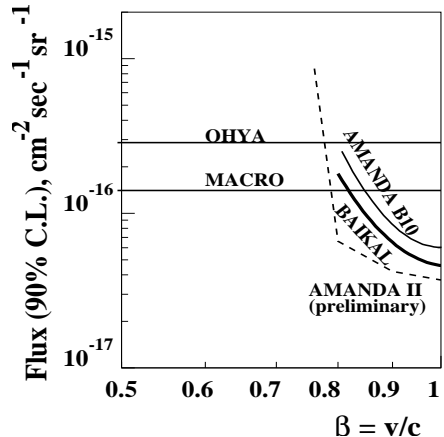}
\parbox[t]{0.47\textwidth}
{\vspace*{-0.2cm}
\caption{Upper limits on the flux of fast monopoles obtained 
in this analysis (Baikal) and in other experiments.}\label{fig6}}
\end{figure}

\subsection{A search for fast magnetic monopoles}

Fast magnetic monopoles with Dirac charge $g=68.5 e$ are 
interesting objects to search for with deep underwater neutrino telescopes.
The intensity of monopole Cherenkov radiation 
(for $\beta$=$1$)
is $\approx$ 8300
times higher than that of muons. 
Optical modules of the 
Baikal experiment can detect such an object from a distance of up 
to hundred meters.  
The processing chain for fast monopoles starts with the selection of
events with a high multiplicity of hit channels: 
$N_{\mbox{\small hit}} > 30$. 
Because of the high background of downward atmospheric muons, 
we restrict 
the search 
to 
upward moving 
monopoles. 
For an upward going particle, the time tags of hit channels increase 
with rising $z$-coordinates from bottom to top of the detector. 
To suppress downward  moving particles, a cut on the value of the 
time--$z$--correlation, $C_{tz}$, is applied.

Within 1003 days of live time 
used
in this analysis, 
about $3.3 \cdot 10^8$ events with 
$N_{\mbox{\small hit}} \ge 4$
have been recorded, with 20943 of them satisfying cut 0 
($N_{\mbox{\small hit}}> 30$ 
and 
$C_{tz}>0$). 
For further background suppression 
(see \cite{APP_MONOP} for details of the analysis)
we use additional cuts, which essentially reject muon events 
and at the same time only slightly reduce
the effective area for relativistic monopoles.

The upper limit on a flux of magnetic monopoles with
$ \beta=1$ is 4.6$\cdot$10$^{-17}$ cm$^{-2}$s$^{-1}$sr$^{-1}$.
In Fig.~\ref{fig6} we compare our upper limit 
for an isotropic flux of fast monopoles 
obtained with the Baikal neutrino telescope 
to the limits from the underground 
experiments Ohya \cite{ohya} and MACRO \cite{macro}, 
and from the underice detectors 
AMANDA-B10 \cite{AMANDAMON} and 
AMANDA-II \cite{AMANDAMON2}(preliminary).

We mention, that 
with an early NT200 prototype,
a search for slow monopoles 
has been published \cite{APP1}
($\beta$=10$^{-5}$--$10^{-3}$ 
and for monopole catalyzed proton decay).
An NT200-analysis, with improved sensitivity down to 
$10^{-17}$cm$^{-2}\cdot$s$^{-1}\cdot$sr$^{-1}$, 
is in preparation.

\begin{figure}
\includegraphics[width=0.45\textwidth, trim= 0cm 0.9cm 0cm 0cm]{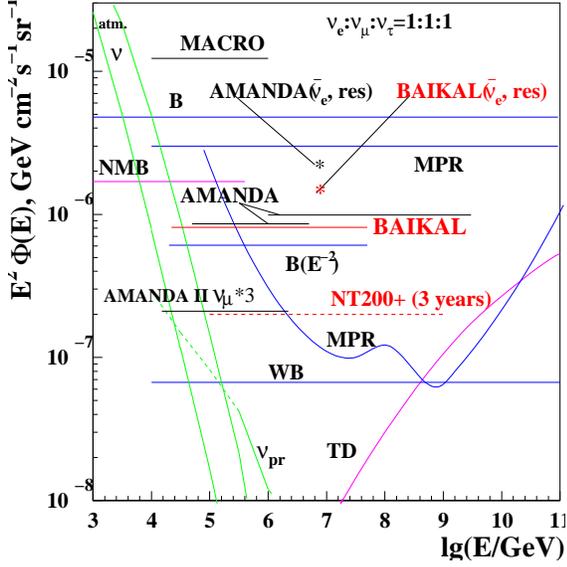}
\\
\caption{
All-flavor neutrino flux 
predictions in different models of neutrino sources 
compared to experimental upper limits to $E^{-2}$ fluxes obtained
by this analysis and other experiments (see text).
Also shown is the sensitivity expected for 3 live years of the
new telescope NT200+ \cite{RW,NT+2}.
}
\label{fig7}
\end{figure}

\subsection{A search for extraterrestrial high-energy neutrinos}

The BAIKAL survey for high energy neutrinos searches
for bright cascades produced at the neutrino interaction
vertex in a large volume around the neutrino telescope \cite{HEB}.
We select events with high multiplicity of hit channels $N_{\mbox{\small hit}}$,
corresponding to bright cascades. 
To separate high-energy neutrino events
from background events, a cut
to select events with upward moving light signals has been developed.
We define for each event
$t_{\mbox{\footnotesize min}}=\mbox{min}(t_i-t_j)$,
where $t_i, \, t_j$ are the arrival times at channels $i,j$ 
on each string, and the minimum over all strings is calculated.
Positive and negative values of $t_{\mbox{\footnotesize min}}$ correspond to 
upward and downward propagation of light, respectively.

Within the 1038 days of the detector live time,
$3.45 \times 10^8$ events 
with $N_{\mbox{\small hit}} \ge 4$ have been recorded. 
For this analysis we used 22597 events with hit channel multiplicity
$N_{\mbox{\small hit}}>$15 and  
$t_{\mbox{\footnotesize min}}>$-10 ns.
We conclude that 
data are consistent with simulated background
for both $t_{\mbox{\footnotesize min}}$ and $N_{\mbox{\small hit}}$ 
distributions. No statistically significant excess above the background 
from atmospheric muons has been observed. 
To maximize the sensitivity to a neutrino signal we introduce a cut in the 
($t_{\mbox{\footnotesize min}},N_{\mbox{\small hit}}$) phase space.

\begin{table}[htb]
\caption{
Expected number of events $N_m$ and experimental model 
rejection factors for astrophysical neutrino source models.} 
\label{tab1}
  \begin{tabular}{@{}lcc|c}
\hline
 & \multicolumn{2}{c|}{BAIKAL \cite{HEB}}  & AMANDA \cite{AMANDAHE,AMANDAMU,AMANDAMU2}\\
\hline
Model & $\nu_e+\nu_{\mu}+\nu_{\tau}$ & $n_{90\%}/N_{\mbox{\footnotesize m}}$ & 
$n_{90\%}/N_{\mbox{\footnotesize m}}$  \\
\hline
  10$^{-6}\times E^{-2}$ & 3.08 & 0.81 & 0.22  \\
  SS Quasar \cite{SS} & 10.00 & 0.25 & 0.21  \\
  SS05 Quasar \cite{SS05} & 1.00 & 2.5 & 1.6  \\
  SP u  \cite{SP}& 40.18 & 0.062 & 0.054  \\
  SP l \cite{SP}& 6.75 & 0.37 & 0.28  \\
  P $p\gamma$ \cite{P}& 2.19 & 1.14 & 1.99  \\
  M $pp+p\gamma$ \cite{M} & 0.86 & 2.86 & 1.19  \\
  MPR \cite{MPR}& 0.63 & 4.0 & 2.0  \\
  SeSi \cite{SeSi} & 1.18 & 2.12 & -  \\
  \hline
\end{tabular} 
\end{table}

Since no events have been observed which pass the final cuts, 
upper limits on the diffuse flux of extraterrestrial 
neutrinos are calculated. For a 90\% confidence level an upper limit 
on the number of signal events of $n_{90\%}=$2.5  is obtained 
assuming an uncertainty in signal detection of 24\% 
and a background of zero events.

A model of astrophysical neutrino sources, for which the total number
of expected events, $N_m$, is larger than 
$n_{90\%}$, is ruled out at 90\% CL. 
Table \ref{tab1} represents event rates and model rejection factors (MRF) 
$n_{90\%}/N_m$ 
for models of astrophysical neutrino sources 
obtained from our search, as well as 
model rejection factors obtained recently
by the AMANDA collaboration \cite{AMANDAHE,AMANDAMU,AMANDAMU2}.

For an $E^{-2}$ behaviour of the neutrino spectrum and a flavor ratio 
$\nu_e:\nu_{\mu}:\nu_{\tau}=1:1:1$, the 90\% C.L. upper limit on the 
neutrino flux of all flavors obtained with the Baikal neutrino telescope  
NT200 is:
\begin{equation}
E^2\Phi<8.1 \times 10^{-7} 
\mbox{cm}^{-2}\mbox{s}^{-1}\mbox{sr}^{-1}\mbox{GeV} ,
\label{eq2}
\end{equation}
for $20\,$TeV$\,<\,$E$_{\nu}\,<50\,$PeV.
For the resonant process 
at the resonance neutrino energy  
$E_0=6.3\times 10^6 \,$GeV the model-independent limit on $\bar{\nu_e}$ is: 
\begin{equation}
\Phi_{\bar{\nu_e}} < 3.3 \times 10^{-20}
\mbox{cm}^{-2}\mbox{s}^{-1}\mbox{sr}^{-1}\mbox{GeV}^{-1}.
\label{eq3}
\end{equation}

Fig.\ref{fig7} 
shows our upper limit on 
the all flavor $E^{-2}$ diffuse flux (eq.\ref{eq2})
as well as the model independent limit on the resonant $\bar{\nu}_e$ flux 
('*') 
(eq.\ref{eq3}). Also shown are the limits obtained by AMANDA 
\cite{AMANDAHE,AMANDAMU,AMANDAMU2}
and MACRO \cite{MACROHE}, theoretical bounds obtained by 
Berezinsky (model independent (B) \cite{Ber3} and for an $E^{-2}$ shape
of the neutrino spectrum (B($E^{-2}$)) 
\cite{Ber4}, by Waxman and Bahcall (WB) \cite{WB1}, by Mannheim et al.(MPR) 
\cite{MPR}, predictions for neutrino fluxes from topological defects (TD) 
\cite{SeSi}, prediction on diffuse flux from AGNs according to Nellen et al. 
(NMB) \cite{NMB}, as well as the atmospheric conventional 
neutrino \mbox{fluxes \cite{VOL}} from horizontal and vertical 
directions ( ($\nu$) upper and lower curves, respectively) and atmospheric 
prompt neutrino fluxes ($\nu_{pr}$) 
from
Volkova et al.\cite{VPPROMPT}.

\begin{figure}[t]
\begin{center}{
\epsfig{file=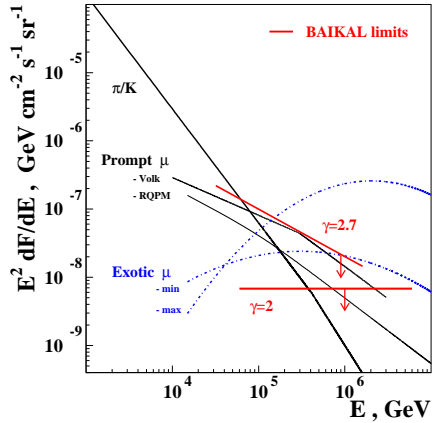,width=0.8\linewidth}
}
\end{center}
\caption{Obtained upper limits on the 
high energy atmospheric muon flux 
(curves with arrows for $\gamma=2$; $2.7$); and
predicted fluxes (for $\pi$/K, prompt; and for exotic).}    
\label{fig_highmu}
\end{figure}

\subsection{A search for high energy muons}

To search for high energy muons,
we use the same event sample and cuts as 
for the diffuse high energy neutrino 
search (see above; 
analysis restricted to 508 live days), 
to which atmospheric muons in turn are a background.
With no events left at final cut level, 
upper flux limits can be derived for 
various muon energy spectra.
Fig.\ref{fig_highmu} shows the experimental limit for a typical 
prompt muon spectrum with $\gamma=2.7$, 
together with theoretical predictions \cite{NE05,VPPROMPT,BUG}.

We can also test for an ``exotic'' component
of high energy atmospheric muons, 
which had been postulated
to explain the ``knee'' in the cosmic ray energy spectrum 
by a new interaction at PeV-scale 
(see \cite{ECRS04,HIGHEMU,HOERANDEL} and ref. therein). 
For such hard 
exotic spectra
(dashed curves in Fig.\ref{fig_highmu}), 
NT200 has a high sensitivity.
Our experimental limit \cite{HIGHEMU}
for a generic $E^{-2}$-spectrum is given in Fig.\ref{fig_highmu},
demonstrating the sensitivity to exclude even the 
lowest exotic flux.

%
\begin{figure*}
\begin{minipage}[b]{.23\linewidth}
\centering\epsfig{figure=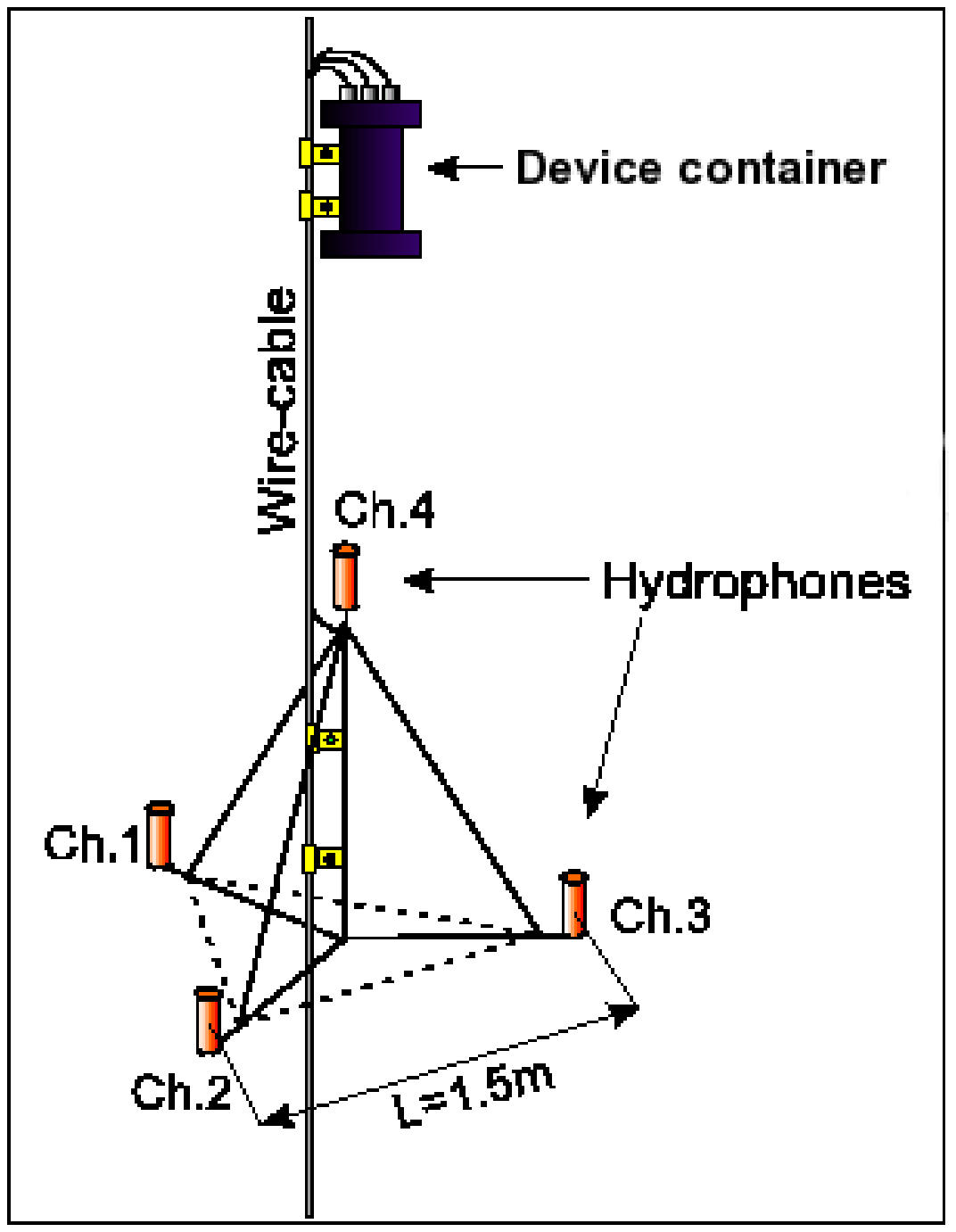,width=0.9\linewidth}
\caption{The stationary 4-channel acoustic device, operating in Lake Baikal.}
\label{fig_acdevice}
\end{minipage}
\hfill  
\begin{minipage}[b]{.35\linewidth}
\centering\epsfig{figure=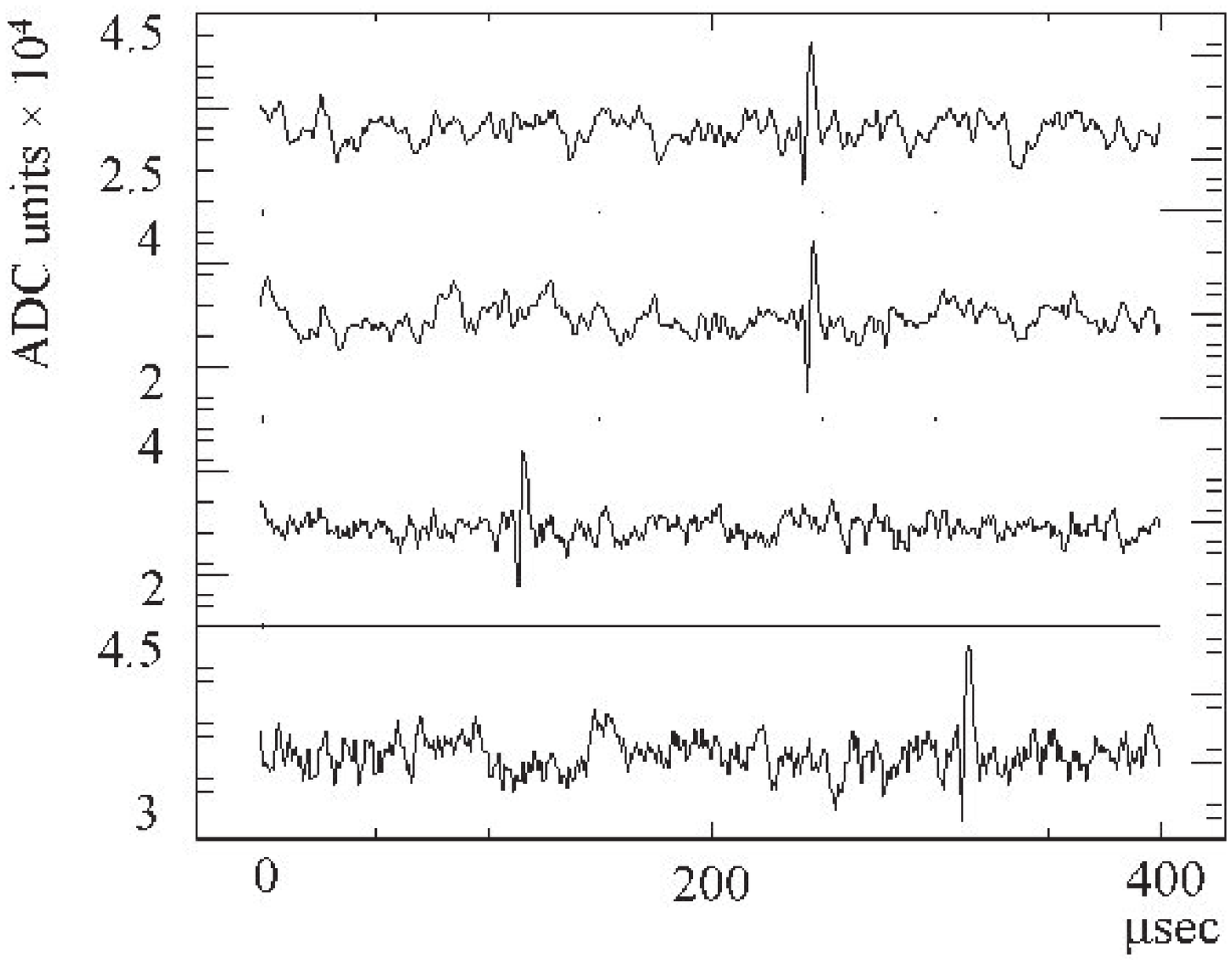,width=0.99\linewidth}
\caption{A triggered event with bipolar pulses (channels $1$--$4$).}
\label{fig_bippulse}
\end{minipage}
\hfill     
\begin{minipage}[b]{.30\linewidth}
\centering\epsfig{figure=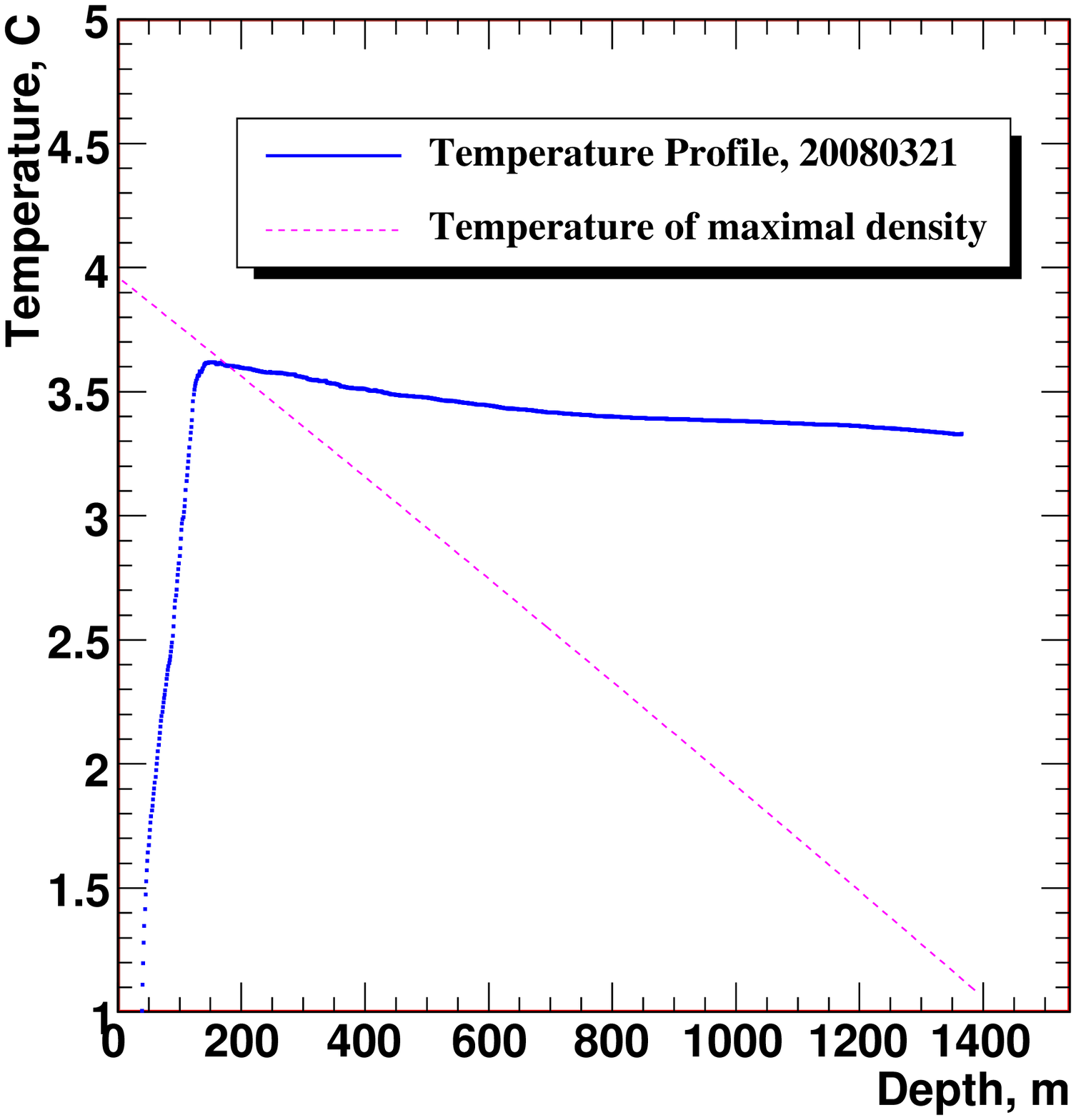, width=0.9\linewidth}
\caption{
Measured temperature as function of depth (28.3.2008) 
and temperature of maximal density, TMD.}
\label{fig_temp2803}
\end{minipage}
\end{figure*}

\section{A prototype device for acoustic neutrino detection in Lake Baikal}
\label{sect_acoustic}

To detect UHE astrophysical neutrinos 
(E$_{\nu}$$>10^{17}$eV), 
detectors with sensitive volumes much beyond km3-scale are needed because 
of the steeply falling energy spectra.
Acoustic detection technology, 
aiming at registration
of the acoustic bipolar signals emitted by 
UHE-cascades, 
is a promising technique - 
and has been 
investigated since several years in Lake Baikal
\cite{ACOUSTIC1,ACOUSTIC2}.
Particularly interesting for our site is the 
very large acoustic absorption length 
for $\sim$$30$\,kHz 
(the peak frequency of acoustic 
signals),
which for freshwater is in the km-range,
almost 100\,times larger than that of Cherenkov radiation
\cite{ACOUSTIC2,Clay}.

We performed a series of 
long-term hydro-acoustic measurements 
to quantify the 
ambient
acoustic noise,
which constitutes the main background for acoustic neutrino detection.
We found,
at stationary and homogeneous meteorological conditions, that 
the integral noise power in the relevant frequency range $20$-$50$ kHz 
reaches 
levels as low as 
$\sim$$1$\,mPa \cite{ACOUSTIC1,ARENA08},
one of the lowest levels measured at currently considered 
acoustic neutrino detector sites.

Extraction of small signals from background requires 
an antenna,
consisting of a set of hydrophones. 
We have constructed a digital hydro-acoustic device with four 
input channels shown in Fig.~\ref{fig_acdevice},
for stationary common operation with 
NT200+. It has been installed in April 2006 at one of the 
telescope moorings at 150\,m depth \cite{ACOUSTIC2}.

Fig.\ref{fig_bippulse} presents 
an event with bipolar pulses in all 4 channels,
detected 
by this device. 
After directional reconstruction,
we find 
that most 
pulses arrive from
the  vicinity of 
the horizontal plane 
\cite{ACOUSTIC2}. 
From the region within $45^\circ$ around the opposite zenith,
i.e. from the deep lake zone below the device,
no events with bipolar pulse form have been observed.

We note, that the water temperature at depths below
400\,m is very stable around $\sim$3.4-3.6$^o$C,
see Fig.\ref{fig_temp2803} for a spring 2008 measurement  
and Sect.\ref{sect_sitestudy} for more details.
As also shown in Fig.\ref{fig_temp2803}, the temperature
is only  equal to that of maximal density
(TMD - dashed curve \cite{Chen}) at shallow depths below $\sim$$200$\,m;
at larger depths they differ significantly, 
since TMD falls steeply 
(by $\sim$0.2$^o$C per $100$\,m depth).

\section{Towards a km3 detector in Lake Baikal: the new technology string}
\label{sect_km3}

The Baikal collaboration has
followed over several years an R\&D program 
for an km3-scale neutrino telescope in Lake Baikal.
The construction of NT200$+$ was a first step 
in this direction.
The existing NT200$+$ is a natural 
laboratory  
to verify 
many new
key elements and design principles
of the 
new telescope.

A Baikal km3-detector 
could be made of building blocks similar to NT200$+$, 
but with NT200 replaced by a single string, still allowing 
separation of high-energy neutrino induced cascades from background 
\cite{RW}. 
It will contain a total of $1300$--$1700$ optical modules (OMs), 
arranged at $90$--$100$ strings with $12$--$16$ OMs each, and 
an instrumented  length of $300$--$350$m. 
Interstring distances will be $\sim$$100$\,m. 
The effective volume for 
cascade events
above $100$ TeV is $0.5$--$0.8$\,km$^3$, 
the threshold for muons is $20$--$50$\,TeV.

The most recent km3-milestone
was the construction and installation of a 
``new technology'' prototype string in spring 2008. 
This string is operating as an integral part of NT200+.
Prototype string design and first results 
are described in detail in these proceedings \cite{Toulon_string}.
It is based on several new technology elements: (1) large area 
hemispherical PMTs (12" Photonis and 13" Hamamatsu, first
time used in an underwater telescope), (2) 200MHz FADC 
readout technology, combined with string-based triggering 
and time synchronization by array trigger time-stamps,
and (3) new calibration 
elements.
Data collection \& transmission is based on 
copper cables using proven DSL-technology (as in NT200+ \cite{RW}).

First calibration and verification tests have been successful, 
see \cite{Toulon_string}.
The string is now 
running in either of 
two operation modes: standalone 
atmospheric muon trigger, 
or in coincidence mode with NT200+ (high energy cascades).
A detailed verification of the prototype 
string will be based on this 
large statistics data sample.

MC-optimization for the km3-detector design is going on, 
as well as studies for optimal trigger technologies. 
A technical design report 
for the new telescope is due for fall 2008.

\section{Related science with the Baikal telescope}
\label{sect_sitestudy}

Related science regarding 
water parameters, 
water exchange and deep-water renewal 
processes \cite{Weiss_Nature} 
are an integral part of the neutrino project.
Optical properties (absorption, scattering) are 
permanentely monitored by custom-made devices \cite{APP1,OPT1} 
while the 
background light recorded 
by the telescope's 3-dimensional light sensor array
contains
rich information on
seasonal changes of water luminescense,
and even allows to 
trace short term flows of 
luminescent matter 
\cite{PROGR98,PYL93}.
The telescope's modern infrastructure
offers 
a natural
opportunity  
to add realtime sensor systems (e.g. seismic, temperature),
either connected by cables or underwater acoustic links  
\cite{KEBKAL}.

The lake's temperature profile
is 
a driving factor 
for  
hydrophysical as well as
hydrobiological and hydrochemical processes;
and at the same time the 
water temperature is a very efficient tracer to 
study 
hydrophysical phenomena in this natural basin. 
Stationary long-term high-precision temperature 
measurements have been carried out since March 1999
in cooperation with the Swiss Federal Institute of 
Environmental Science and Technology (EAWAG). 
More than 50 temperature loggers,
distributed between the lake bottom and 
$15$ m 
depth at $3$ moorings in the vicinity of NT200+,
record the 3-dimensional temperature profile all year round at $10$ min intervals 
with a resolution better than $0.002^\circ$C
\cite{Schmidt,NOVOSIB08}.

\begin{figure}[t]
\vspace*{-0.3cm}
\centering
\includegraphics[width=0.47\textwidth, trim= 0cm 0cm 0cm 0cm]{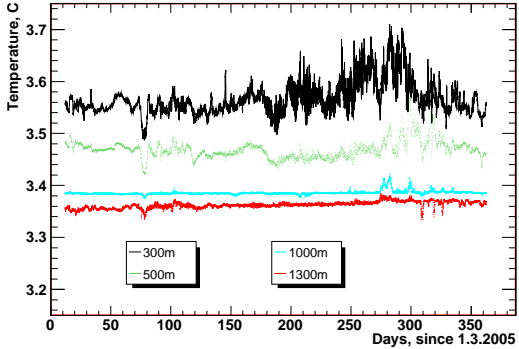}
\vspace*{-0.7cm}
\caption{
Vertical temperature profile
from 300\,m (top curve) to 1300\,m depth (bottom curve; $\sim$50\,m above lake-bed), 
as measured during 2005.}
\label{fig_deeptemp}
\end{figure}

The largest seasonal temperature variations occur at 
shallow depths $<$200\,m,  
as discussed in \cite{NOVOSIB08}.
Temperatures measured at depths $300$--$1300$\,m 
in 2005 are, as an example, shown in Fig.\ref{fig_deeptemp}.
The 
temperature below $300$\,m slowly decreases with depth
and 
the 
density of water is perfectly stratified all around the year.
Almost simultaneous temperature variations down to $1000$\,m depth are observed. 
The structures of these variations at different depths clearly correlate; 
their 
amplitude decreases with depth \cite{NOVOSIB08}.
These disturbances have a maximum magnitude 
close to the
autumn homothermy and are connected with intrusion of water from
upper layers.

\begin{figure}[tb]
\centering
\includegraphics[width=0.27\textwidth] {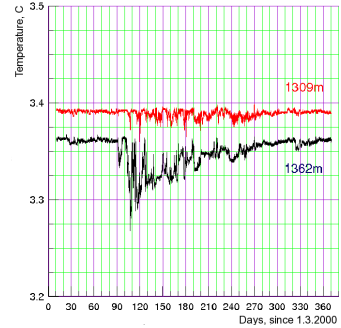}
\vspace*{-0.2cm}
\caption{Temperature at 
$1309$\,m and $1362$\,m depth 
($4$\,m and $57$\,m above 
bottom), measured from March, 2000 - March, 2001. 
For the huge cold-water intrusion in June, see text.} 
\label{fig_intrusion}
\end{figure}

However, regulary observed 
huge cold water intrusions into the 
near-bottom zone
(occuring around the time of convective surface mixing in June and December/January),
which have a typical volume of about $10$ km$^3$,
could not be explained by this vertical water exchange,
since data 
did not show 
vertical disturbances to reach 
down to the bottom with signifcant amplitude. 
These intrusions play a 
crucial role in oxygenation of deepest layers 
and lead to an efficient recycling of the nutrients from the deep water. 
But up to now, a convincing explanation of this phenomenon 
had been missing.
One of the largest 
such 
intrusions 
was observed in June 2000, 
when the temperature at 
$1362$\,m depth (just above the bottom) suddenly decreased 
by almost
$0.1$ $^\circ$C and returned to its initial value only 
after 6 months (see Fig.~\ref{fig_intrusion}). 
It has been 
the
3-dimensional time-resolved temperature 
monitoring, 
that 
gave the first experimental
proof that these advective 
intrusions are caused by coastal downwelling and subsequent thermobaric 
instability along the steep lake shores 
- thus solving a longstanding mystery \cite{Schmidt,NOVOSIB08}.

\section{Conclusion}

The Baikal neutrino telescope NT200 has been taking data since April 1998. 
The upper limit obtained for a diffuse 
astrophysical 
($\nu_e+\nu_{\mu}+\nu_{\tau}$) 
$E^{-2}$-flux 
is $E^2 \Phi = 8.1 \times 10^{-7}$cm$^{-2}$s$^{-1}$sr$^{-1}$GeV.
The limits on the flux of fast magnetic monopoles and on 
a 
muon
flux induced by WIMP annihilation at the center of the Earth
belong to the most stringent limits existing to date.  
The limit on a $\bar{\nu_e}$ flux
at the resonant  energy 6.3$\times$10$^6$GeV is presently the most
stringent.
NT200 also has a high sensitivity for exotic UHE atmospheric muon fluxes.

The significantly upgraded telescope NT200+, 
a detector with about 5 Mton enclosed volume, 
has been operating since April 2005,
and has 
an
improved sensitivity for 
a diffuse flux of extraterrestrial neutrinos \cite{RW,NT+2}. 
An ongoing feasibility study for acoustic UHE neutrino detection 
with a stationary antenna 
suggests that favourable conditions exist at Lake Baikal.

Multidisciplinary science activities 
have focused on water monitoring 
(optical absorption/scattering, luminescense) 
and
high precision temperature measurements at 
stationary
moorings over almost a decade. These
have led to 
deeper insight into  
the Lake's ecosystem phenomena.
An upgrade to a 
realtime ecosystem-underwater data network
is feasible.

For the planned km3-scale neutrino detector in Lake Baikal, R\&D-activities 
are in progress.
An important km3-milestone
was the Spring 2008 installation of a 
new technolgy km3-prototype string, 
with full FADC-readout and large area 
hemisperical PMTs (12''/13''),
which now operates together with NT200+.
The km3-detector Technical Design Report is planned for fall 2008.

\section{Acknowledgements}

This work was supported by the Russian Ministry of Education and
Science, the German Ministry of Education and Research, 
REC-17 BAIKAL and the Russian
Fund of Basic Research (grants 08-02-00432, 08-02-10010, 07-02-00791,
08-02-00198, 08-02-10001),
and by the Grant of President of Russia NSh-4580.2006.2. and by 
NATO-Grant NIG-9811707(2005).

\end{document}